\documentclass[pre,twocolumn,showpacs,preprintnumbers,amsmath,amssymb]{revtex4}
\usepackage{graphicx}
\usepackage{epstopdf}

\begin{document}

\title{Delayed feedback as a means of control of noise-induced motion}

\author{N.B. Janson$^{1,2}$, A.G. Balanov$^{1}$, E. Sch\"{o}ll$^1$}

\affiliation{$^1$Institut f{\"u}r Theoretische
Physik, Technische Universit{\"a}t Berlin, Hardenbergstra{\ss}e 36,
D--10623 Berlin, Germany \\
$^2$Department of Mathematical Sciences, Loughborough University, Loughborough, Leicestershire LE11 3TU, UK }

\date{\today}

\begin{abstract}
Time--delayed feedback is exploited for
controlling noise--induced motion in coherence resonance
oscillators. Namely, under the proper choice of time delay, one can either
increase
or decrease the regularity of motion. It is shown that in an excitable
system, delayed
feedback can stabilize the frequency of oscillations against variation of
noise strength. Also, for fixed noise intensity, the phenomenon of
entrainment of the  basic oscillation period by the delayed
feedback occurs. This allows one to steer the timescales of
noise-induced motion by changing the time delay. \end{abstract}


\maketitle

Very often in practical application the need arises to control the
properties of oscillations. Usually control assumes an enhancement in
regularity of motion
\cite{arrhythmia,BAB02,Balanov_CSF03}.
However, in some cases, for instance in medical applications,
one aims to disorder oscillations, since too strong coherence might be
undesirable and even lead to damaging consequences,
e.g. epilepsy or Parkinson's disease,
\cite{epilepsy, Tass}.
During the last decade new methods for control of irregular
self-sustained oscillations
in deterministic systems have been developed,
including suppression of
chaos by an external (periodic) signal \cite{chaos_supp},
stabilization of unstable periodic orbits embedded in a chaotic attractor
by time-discrete control \cite{OGY}, or the use of a time-delayed
feedback loop \cite{Pyragas_PLA92,SOC94} for the same purpose.

Whereas the existing methods are designed to control deterministic
oscillations or, most recently, noise-induced enhancement of deterministic
oscillations \cite{Gammaitoni99, Lindner01} and self-oscillations affected
by noise \cite{Goldobin},
there is a large class of systems
that do not oscillate autonomously; but if they are forced even
by a purely random process featuring no specific timescales, they
demonstrate motion resembling a self-oscillatory process
\cite{CR_Gang,CR_Strogatz}.
Prominent representatives of this class are excitable systems like neurons
\cite{CR_neuron}, chemical reaction systems \cite{Chem},
and semiconductor nanostructures \cite{UNK03}.
The degree of closeness of their oscillations to ideally periodic
ones, i.e. coherence, can depend resonantly on noise intensity
\cite{CR_Gang},
which is why it was called {\em coherence resonance} (CR)
\cite{Pikovsky_PRL97}. CR
has been shown to occur in systems
close to bifurcations \cite{CR_Neiman},
in excitable systems \cite{Pikovsky_PRL97},
and in bistable systems \cite{Lindner00}.
Remarkably, CR oscillators
possess the fundamental property of self-oscillators, namely, the
ability to synchronize \cite{CR_Postnov}.

Frequently the timescale
of oscillations in a CR system varies substantially depending on noise
intensity. Since the latter is not easily
controllable in practice, there is need to make a CR device robust
against variation in the properties of noise. Another important
task is to find a reliable way to deliberately change the timescales
of noise-induced oscillations in a universal way without affecting
intrinsic system parameters. Finally, the obvious need is to control
the regularity of noise-induced motion. At present, all three
problems remain a challenge. In the present Letter we propose to
exploit time-delayed feedback control to tackle all three issues.

As a first example of a CR oscillator, we consider the noisy
Van der Pol system closely before the Hopf bifurcation,
extended by a delayed feedback loop
\begin{eqnarray}
\label{VDP}
\frac{dx}{dt}&=&y, \\
\frac{dy}{dt}&=&(\nu -x^2)y-\omega_0^2x +K(y_{\tau}-y) +D \xi(t).
\nonumber
\end{eqnarray}
Here, $x$ and $y$ denote phase variables at
time $t$, while $y_{\tau}$ denotes the delayed variable $y(t-\tau)$;
$K$ is the strength of delayed feedback. $\xi(t)$ is a random variable
with Gaussian distribution, zero mean and unity variance, $D$ is the noise
intensity. We set the parameters $\nu=-0.01$ and $\omega_0=1$
at which a stable focus exists.

First consider $K=0$, i.e. no delayed feedback in Eq. (\ref{VDP}). While no
limit cycle occurs
at $D=0$, application of noise induces oscillatory motion as
illustrated by the phase portrait in Fig. 1 (a). The coherence of
oscillations may be quantified by the
correlation time $t_{cor}$, estimated from the normalized autocorrelation
function $\Psi(s)$ of $y$ as $t_{cor}=\int_0^{\infty} |\Psi(s)|ds$.
In Fig. 2(a) the grey (green on-line) line shows $t_{cor}$
vs $D$ for $K=0$. 
At small noise intensity D the oscillations are more coherent.

In deterministic self-oscillatory systems, application of a delayed feedback
in the form above acts as follows.
If there exists an unstable periodic orbit of period $T$ in the phase
space, a feedback with delay time $\tau=T$
can stabilize this orbit in some range of the control strength $K$.
A CR system may have no periodic orbits,
but noise may induce oscillations with a well-defined timescale
that is associated with the spectral peak.
We suppose that application of a delayed feedback can act by
analogy with a system containing a periodic orbit, provided that $\tau$ is
equal or close enough to the basic period $T_0$ of the noise-induced
motion \cite{comment1} without feedback. Namely, it
should suppress deviations from a reference state and thus enhance
the regularity of oscillations.

To test this expectation, we switch on the control force in Eq.
(\ref{VDP}).  We set $\tau=T_0$ with $T_0=6.17283951 \approx
2\pi/\omega_0$.
The phase portrait
with control at $K=0.2$ is shown in Fig. 1(b) for the same $D$ as in (a).
Fig. 1(b)
reveals a remarkable ordering of the oscillation as compared with Fig. 1(a).

To quantify the ordering due to the feedback, we estimate
$t_{cor}$ in dependence on $D$ as above. It is given for $K=0.2$ by
the black line in Fig. 2(a). One can see that for any $D$ the coherence of
noise-induced oscillations becomes larger when the delayed feedback
loop is switched on and $\tau$ is close to $T_0$.
On the other hand, it was found that if $\tau$ is far from an integer
multiple of $T_0$, the coherence of oscillations, on the
contrary, decreases. Also, for $\tau=T_0$ the coherence of
noise-induced oscillations in Eq. (\ref{VDP})
was found to increase  monotonically
with increasing $K$.
In the following we fix $K=0.2$.
\begin{figure}
\begin{center}
\includegraphics[width=0.45\textwidth]{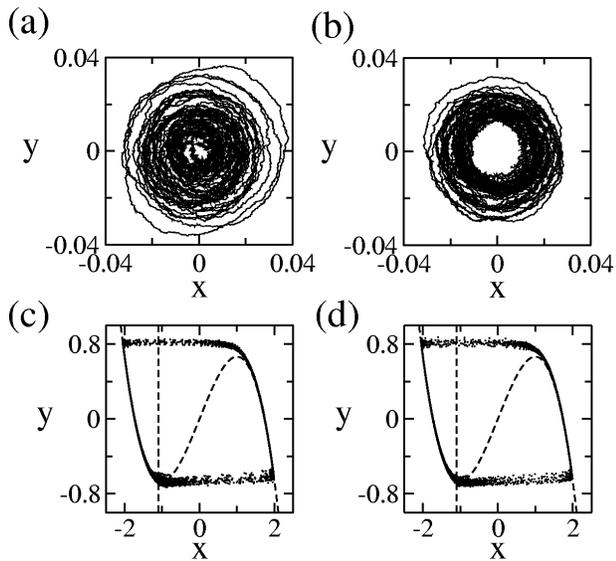}
\caption{Phase portraits of noise-induced motion: (a),(b) Van der Pol
oscillator at $D=0.003$, (c),(d) FitzHugh-Nagumo system at $D=0.09$
(the dashed lines denote the null-isoclines),
(a),(c) $K=0$; (b),(d) $K=0.2$, $\tau=T_0$. }
\end{center}
\end{figure}
\begin{figure}
\begin{center}
\includegraphics[width=0.48\textwidth]{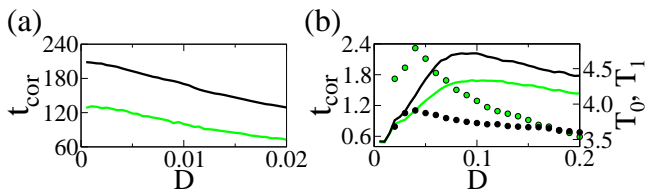}
\caption{
(color on-line)
Correlation time $t_{cor}$ vs. noise intensity $D$ for (a) Van der Pol
oscillator, (b) FitzHugh-Nagumo system. Grey (green on-line) lines: $K=0$, black lines:
$K=0.2$, $\tau=T_0$. (b) grey (green on-line) circles: $T_0$ for $K=0$,
black circles: $T_1$ for $K=0.2$, $\tau=T_0$.}
\end{center}
\end{figure}
Next, we study how the feedback can affect the system's
timescales. For this purpose, we set the noise at the
intensity $D=0.003$ as for Figs. 1(a), (b),
and follow the evolution of Fourier
power spectra with $\tau$, which is illustrated by Fig. 3(a).
Without feedback ($\tau=0$), system (\ref{VDP})
has one pronounced peak $f_0$ in the spectrum.
As $\tau$  increases from zero, the peak frequency, height and width
change. At $\tau>8$, new peaks
that change their positions, heights and widths with $\tau$
become clearly visible.

Since the control parameter of delayed feedback is the time interval
$\tau$,
we propose
to describe the response of the system in terms of
periods rather than frequencies.
With feedback, we select
all spectral peaks, and for each peak introduce the period $T$ as
the inverse of the peak frequency. Denote the period of highest peak by $T_1$.
The dependences of $T$ on $\tau$ are given in Fig.
4 (a):  $T_1$ by circles (yellow on-line), other $T$ by
crosses (blue on-line). One can see that variation of $\tau$ changes
$T_1$ in a certain range, doing so most effectively as
$\tau < T_0/2$. As $\tau$ increases beyond $T_0/2$,
$T_1$ drops quickly, and again increases with $\tau$ with a similar
slope as before. After $\tau$ has increased by about $T_0$,
$T_1$ again drops abruptly to a lower branch, and again follows $\tau$,
although with a smaller slope. These abrupt transitions to successive lower
branches occur roughly every $T_0$ time units, and each subsequent
entrainment happens at a lower slope. The plot of $T_1$ {\it vs} $\tau$
exhibits a piecewise approximately linear dependence, the larger the $\tau$,
the closer each segment is to a straight line.

The numerical results obtained above for the Van der Pol system
can be understood in terms of a general theory of a canonical nonlinear
oscillator with time-delayed feedback
\begin{eqnarray}
\label{can_delay}
\ddot{x}+f(x,\dot{x})-K(\dot{x}_{\tau}-\dot{x})=0.
\end{eqnarray}
Note that  Eq. (\ref{VDP}) fits the form Eq. (\ref{can_delay})
if rewritten as a single second-order differential equation
with $f(x,\dot{x})=-(\nu -x^2)\dot{x}+\omega_0^2x$ and $y=\dot{x}$.
Without feedback ($K=0$ or $\tau=0$) the fixed point $(x_0,0)$ is a stabe focus
if
\begin{equation}
\label{ders}
0<\frac{\partial f}{\partial \dot{x}}<2\sqrt{\frac{\partial
f}{\partial x}},
\end{equation}
where partial derivatives are taken at the fixed point.
In Eq. (\ref{VDP}) $x_0=0$, and Eq. (\ref{ders}) is true for the given
parameters. Setting $K>0$ does not change $x_0$. Either with, or without
feedback
the noise-induced oscillations take place in the close vicinity of the fixed
point. It is to be expected that the motion is influenced by
the local properties of this point.
At $\tau=0$ the stable focus has a pair of
complex conjugate eigenvalues $\lambda_0=p_0 \pm iq_0$, $p_0<0$, $q_0
\ne 0$, and the value of $q_0$ should give an estimate of the angular frequency.
Indeed, the only peak of the power spectrum (Fig. 3(a)) has frequency
$f_0 \approx |q_0|/2\pi$.
With $\tau>0$, the system becomes infinite-dimensional, and
possesses a
countable set of eigenvalues $\lambda$.
In order to exclude that the delayed feedback
might induce the birth of a stable limit cycle via a Hopf bifurcation,
thus providing a trivial explanation for the remarkable ordering of
oscillations, we perform a linear stability analysis of the fixed point
of Eq. (\ref{can_delay}).
Following the standard routine of linearizing Eq. (\ref{can_delay})
around the fixed point \cite{Hale}, the characteristic
equation for $\lambda$ is derived:
\begin{eqnarray}
\lambda^2+\lambda\frac{\partial f}{\partial \dot{x}}+
\frac{\partial f}{\partial x}-K\lambda(e^{-\lambda \tau}-1)=0,
\label{pq-f}
\end{eqnarray}
Substituting $\lambda=p + iq$, real and imaginary parts can be 
separated.
The condition for a Hopf bifurcation is $p=0$, $q \ne 0$.
Substituting it into the imaginary part of Eq. (\ref{pq-f}) we obtain:
\begin{eqnarray}
\label{nozerop}
\cos q\tau =\frac{K+\partial f/\partial \dot{x}}{K}.
\end{eqnarray}
Since the right-hand side is larger than unity due to Eq. (\ref{ders}),
the Hopf bifurcation condition is not satisfied for any $K$ and $\tau$.
Thus, the delayed feedback in the given form {\it cannot} induce a Hopf
bifurcation.

The numerical solution of Eq. (\ref{pq-f}) with $f(x,\dot{x})$ from
Eq. (\ref{VDP}) yields the eigenvalue spectrum
$\lambda=p + iq$ as a function of $\tau$. The eigenperiods defined as
$T^e=2\pi/|q|$ (dots in Fig. 4(a)) coincide
remarkably with the inverse peak frequencies of the power spectrum of the
noise-induced oscillations as a function of $\tau$. The corresponding
real parts $p$ are given by dots in Fig. 4(b) (the seven largest
$p$ are shown).
All $p$ remain negative, but, as seen from Fig. 4(b), nonmonotonically
change with $\tau$. As $\tau$ increases, separate branches of $p$ cross,
thus providing a striking explanation of the strongly nonmonotonic,
discontinuous evolution
of the dominant spectral peak of the noise-induced motion under delayed
feedback: The period $T_1$ of the highest spectral peak
(circles (yellow on-line) in Fig. 4(a)) always coincides with
the period  $T^e$ of the least stable eigenmode, i.e. the one with the
largest real part which we denote as $p_1$ (circles (yellow on-line)
in Fig. 4(b)). The more stable eigenmodes result in the side peaks of
the frequency spectrum. The more stable the modes are, the lower the
peaks are.

As $p_1$ oscillates with $\tau$, the degree of stability of the fixed point
of the deterministic system is modulated,
thus leading to modulation of the coherence
of the stochastic motion, quantified by the correlation time $t_{cor}$
(solid line (green on-line) in the upper part of Fig. 4(b)).
The local maxima of coherence occur when
$p_1$ is close to zero, and $T_1$ is close to $T_0$.

The entrainment of $T_1$ by $\tau$, which manifests itself in the
almost piecewise linear dependence of $T_1$
on $\tau$ for large $\tau$, can be understood as follows.
As shown above, it is related
to the eigenvalue whose real part $p_1$ is closest to zero. Assuming
$p_1 \approx 0$ in Eq. (\ref{pq-f}) we obtain Eq. (\ref{nozerop}).
With $\partial f/\partial \dot{x}=-\nu \ll K$,
and $(K-\nu)/K\approx 1$, this gives $\cos(q \tau) \approx 1$ and
$|q| \tau \approx 2\pi n$, where $n$ is integer.
Then the eigenperiod $T^e$ is
\begin{eqnarray}
\label{linear}
T^e=2\pi/|q| \approx \frac{\tau}{n}.
\end{eqnarray}
As illustrated by Fig. 4(b), $p_1$ is close enough to zero only for
large $\tau$, for which the relation (\ref{linear}) holds most
accurately. To obtain the location of the maxima of $p_1$, i.e. the
maxima of coherence, substitute $p_1 \approx 0$, $q = 2\pi n/\tau$
into the real part of Eq. (\ref{pq-f}), which yields
$\tau = 2\pi n/\omega_0=n T_0$.

To summarize, delayed feedback applied to an oscillatory system of the form
(\ref{can_delay}), gives rise to a countable set of eigenmodes of the
fixed point, whose
eigenperiods and stability are controlled by $\tau$.
The highest peak in the spectrum of the noise-induced motion is due to
excitation of the least stable eigenmode. The coherence of oscillations is
the higher, the less stable the mode is. The range of modulation of
the peak frequency is largest for small tau and large K.

\begin{figure}
\begin{center}
\includegraphics[width=0.48\textwidth]{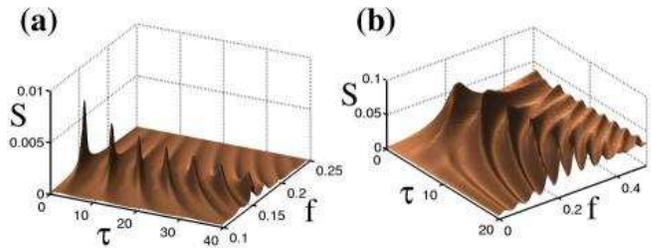}
\caption{Fourier power spectra of noise-induced oscillations in
dependence on $\tau$ for (a)
Van der Pol oscillator at $D=0.003$; (b) FitzHugh-Nagumo system at
$D=0.09$, $K=0.2$. The spectrum is computed from $y$.
}
\end{center} \end{figure}

\begin{figure}
\begin{center}
\includegraphics[width=0.48\textwidth]{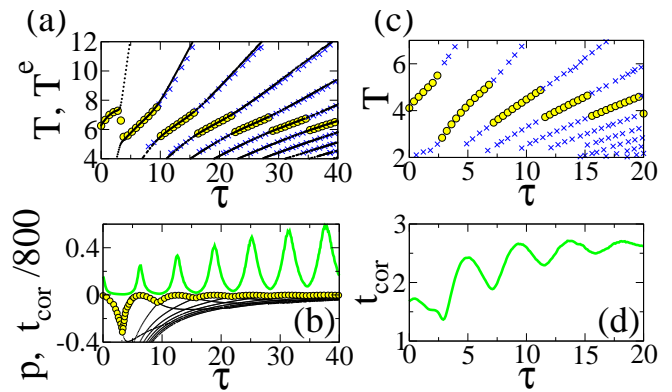}
\caption{(color on-line) Spectral peaks, coherence and eigenvalues vs
$\tau$ at $K=0.2$.
(a),(b): Van der Pol oscillator at $D=0.003$,
(c),(d): FitzHugh-Nagumo system at $D=0.09$.
(a),(c) crosses (blue on-line): $T$,  circles (yellow on-line):
$T_1$, black dots: $T^e$.
(b),(d) solid line (green on-line):
$t_{cor}$, (b) black dots: seven largest $p$, circles (yellow
on-line): $p_1$. }
\end{center}
\end{figure}

Next, we consider another example of a CR oscillator, the FitzHugh-Nagumo
system, which serves as a prototype of an excitable system.
Extending it again by a delayed feedback loop, we obtain
\begin{eqnarray}
\label{FHN}
\epsilon \frac{dx}{dt}&=&x-\frac{x^3}{3}-y, \\
\frac{dy}{dt}&=&x+a+K(y_{\tau}-y)+D \xi(t). \nonumber
\end{eqnarray}
We set the parameters
$\epsilon=0.01$ and $a=1.1$ such that a stable node is the only
attractor of the system in the absence of feedback.

Without feedback ($K=0$), the mechanism for
inducing oscillations by noise is different from the Van der Pol oscillator
(\ref{VDP}). In Fig. 1(c), (d) dashed lines show the null-clines
defined by $dy/dt=0$ (vertical) and by $dx/dt=0$ (cubic
parabola). They intersect at
the fixed point, which is slightly displaced
to the left of the minimum of the parabola for the parameters chosen.
The null-clines divide the phase plane into four regions
with different directions of phase velocity. Dots in Fig. 1(c)
show the phase portrait with noise $D=0.09$.

In Fig. 2(b) the grey (green on-line) line shows $t_{cor}$ for Eq. (\ref{FHN})
versus noise intensity, exhibiting a distinct maximum at $D=0.09$.
Also, grey (green on-line) circles show the basic period $T_0$ of oscillations.
Unlike Eq. (\ref{VDP}), here $T_0$ changes substantially with noise,
as was earlier shown in \cite{Lindner00}.

Now, switch on the feedback with $K=0.2$ and set $\tau$ equal to the value
of
$T_0=4.12694$ at optimum noise. The black line in Fig. 2(b) denotes $t_{cor}$
vs $D$,
and shows that for any $D$ the coherence of oscillations is higher
with the feedback. However, this feature
is not visible in the phase portrait (Fig.
1(d)). Black circles in Fig. 2(b) show the basic period $T_1$
with feedback. It is evident that delayed feedback substantially
reduces the variation
of the noise-induced basic timescale. Note, however, that
this may not be so if $\tau$ is very different from $T_0$.

Next, we study how the feedback can control the timescales
and the regularity of noise-induced motion. Fix $D$
at an optimum value $0.09$, $K$ at $0.2$ and change $\tau$.
The spectrum in dependence on $\tau$ is given in Fig. 3(b).
With increasing $\tau$, the spectral peaks move towards zero, and
the spectrum is gradually enriched by new peaks.
As with the Van der Pol oscillator, select all visible peaks
with periods $T$,
and denote the period of the highest peak as $T_1$.
In Fig. 4(c)
$T$ of several peaks are given by crosses (blue on-line), and
$T_1$ by circles (yellow on-line), depending on $\tau$.
These dependencies are qualitatively very much like those for Eq.
(\ref{VDP}) in Fig. 4(a).  In Fig. 4(d) $t_{cor}$ is given
depending on $\tau$, exhibiting oscillatory features. Local maxima of
coherence occur when $T_1$ is equal to $T_0$.
Unlike in case of the Van der Pol oscillator, the mechanism of delayed
feedback control can not be explained by a local analysis of the fixed point
since the oscillations are characterized by large excursions in phase space.
Rather, a global analysis would be needed which is clearly beyond the scope
of the present Letter.

In conclusion, time-delayed feedback in the form of the difference
between the current and a delayed state of the system can be used
to control oscillations that are induced merely by noise.
The most crucial parameter of such a control is the time delay $\tau$,
depending on which the coherence of noise-induced oscillations
increases or decreases. With this, a phenomenon of entrainment of the basic
period of noise-induced motion by the time delayed feedback is discovered.
The latter ability is somehow reminiscent of classical
synchronization phenomena, in that the externally imposed timescale $\tau$
tunes the basic period of oscillations in the system, although it involves
quite different mechanisms.

This work was supported by DFG in the framework of Sfb 555.
The authors gratefully acknowledge discussions with A. Nikitin and A. Amann. 


\begin{thebibliography}{99}
\bibitem{comment1} By basic period we mean the inverse of the
frequency of the highest spectral peak.
\end{thebibliography}

\end{document}